\documentclass[a4paper, 11pt]{article}
\usepackage{helvet}
\usepackage{amsmath}
\usepackage{amsfonts}
\usepackage{amssymb}
\usepackage[english]{babel}
\usepackage{graphicx}
\usepackage{amsthm}
\usepackage{epstopdf}
\usepackage{tikz}
\usepackage[title]{appendix}

\usepackage{longtable}
\usepackage{tabularx}
\usepackage{multirow}
\usepackage{float}
\allowdisplaybreaks

\newtheorem{theorem}{Theorem}[section]
\newtheorem{definition}{Definition}[section]
\newtheorem{lemma}[theorem]{Lemma}

\newtheorem{conjecture}[theorem]{Conjecture}

\def\whitebox{{\hbox{\hskip 1pt
 \vrule height 6pt depth 1.5pt
 \lower 1.5pt\vbox to 7.5pt{\hrule width
    3.2pt\vfill\hrule width 3.2pt}%
 \vrule height 6pt depth 1.5pt
 \hskip 1pt } }}
\def\qed{\ifhmode\allowbreak\else\nobreak\fi\hfill\quad\nobreak
     \whitebox\medbreak}

\newcommand{\ignore}[1]{}
\usepackage[numbers,sort&compress]{natbib}

\begin{document}
\baselineskip 16pt

\title{Bounds and Constructions for Optimal Wide-Gap Frequency Hopping Sequences}

\author
{
\footnotesize \ Jiwen Zhang and Haitao Cao\thanks{E-mail: caohaitao@njnu.edu.cn}\\
      \scriptsize     School of Mathematical Sciences, Ministry of Education Key Laboratory for NSLSCS,\\ \scriptsize Nanjing Normal University, Nanjing 210023, China.\\
}

\date{}
\maketitle
\begin{abstract}
In a frequency-hopping (FH) system, to reduce interference to the received signal, each user's frequency hopping sequence (FHS) shall have wide-gap property.
In this paper, we update the upper bounds on the minimum gaps of optimal uniform wide-gap frequency hopping sequences (WGFHS) with parameters $(v,s,\lambda;e)$ where $v=ks, k\ge 3, \lambda=k$ and $s\ge k(k-1)+1$ is odd.
We completely prove the existence of such WGFHSs for $k\le 3$.


\smallskip
\end{abstract}

\section{Introduction}

As a fundamental spread-spectrum technique, frequency-hopping (FH) communication offers robustness against fading, interference, and interception \cite{SOSL}. It is widely used in both civilian and military applications, including Bluetooth, HomeRF, ultra-wideband (UWB) communications, smart grids, GSM, and radar communications \cite{GEK,LB,ONS,S2002,STH,ZLP}. When multiple users simultaneously share the same frequency band, frequency collisions may occur, leading to interference characterized by the Hamming correlation functions of the FH sequences \cite{S1994}. Such interference can cause demodulation errors at the receiver, thereby degrading communication performance. To establish theoretical limits on Hamming correlation, Lempel and Greenberger \cite{LG} and Peng and Fan \cite{PF} derived two fundamental lower bounds.

Let $F=\{f_0,f_1,\ldots,f_{s-1}\}$, where $f_i$ is a frequency, and $s$ is a positive integer. Generally, $F$ is defined by a set of integers $\{0,1,\ldots,s-1\}$.
Let $S$ denote an FHS set of period $v$ over $F$. The Hamming cross-correlation of two FHSs $X={(x_k)}_{k=0}^{v-1}, Y={(y_k)}_{k=0}^{v-1}\in S$ at a shift $\tau$ is defined by
$$H_{X,Y}(\tau) = \sum_{k=0}^{v-1}h[x_k,y_{k+\tau}], \quad 0\le\tau<v,$$
where $h[a,b]=0$ if $a\neq b$ and 1 otherwise. In particular, if $X=Y$, then $H_{X,X}$ is called the Hamming autocorrelation, and the maximal Hamming autocorrelation is defined by
$$ \lambda=H(X) = \max_{1\le\tau<v} H_{X,X}(\tau). $$
	
Moreover, the dynamic nature of interference in modern communication systems makes signals with prolonged frequency dwell times susceptible to tracking interference, motivating the use of shorter dwell times.
In addition to optimal Hamming correlation, FH systems may require a wide-gap property, whereby adjacent hopping frequencies are separated by more than a prescribed positive integer. This requirement has motivated the study of WGFHSs \cite{M1994,WZSS,C1983,MZ}. Lower bounds on the Hamming correlation of WGFHSs have been further investigated in \cite{LFY}. Compared with conventional FHSs, relatively few constructions of WGFHSs have been reported \cite{C1983,HF,HZ,LFY,LFM,MY,MZ,M1994,WW,RW,LLZL,WH,WNW}. Related extensions include wide-gap-zone FHSs \cite{S2024} and WGFHSs with no-hit zones \cite{Z2025} or low-hit zones \cite{Y2025, W2026,Y2026}.
The gap is defined as follows. For an integer $z$, let $\langle z\rangle_{m}$ be its least nonnegative residue modulo $m$.

\begin{definition}
For any given FHS $X={(x_j)}_{j=0}^{v-1}$ over a frequency set $F=\{0,1,\dots,s-1\}$, define
	$e=\min\left\{|x_{\langle j+1\rangle_{v}}-x_j|-1:j=0,1,\dots, v-1\right\}$.
If $e>0$, then $X$ is called a WGFHS with minimum gap $e$.
 \end{definition}

For brevity, a WGFHS with parameters $(v,s,\lambda;e)$ is referred to as a WGFHS of period $v$ over a frequency set $F$ of size $s$, with minimum gap $e$ and the maximal Hamming autocorrelation $\lambda$.
A WGFHS with parameters $(ks,s,\lambda;e)$ where each frequency appears $k$ times is called {\it uniform}.
The Lempel-Greenberger bound for WGFHSs was improved in \cite{LFY}.

\begin{theorem}(\cite{LFY})\label{bound}
Let $X$ be an WGFHS of period $v$ over $F$ of size $s$. Then
\begin{align}
H(X) \geq\left\lceil\frac{(v-\varepsilon)(v+\varepsilon-s)}{s(v-3)}\right\rceil,
\end{align}
where $\varepsilon=v-\left\lfloor\frac{v}{s}\right\rfloor \cdot s.$ If $X$ achieves the equal sign of Eq.(1), then $X$ is said to be optimal.	
\end{theorem}

This paper focuses on uniform WGFHSs with $k\ge2$ and $s\geq 5$. For these sequences, the optimality condition is exactly $\lambda=k$, and the upper bound of the minimum gap is $e\leq \lfloor \frac{s-1}{2} \rfloor -1$ which is presented in \cite{LLZL}.
In Section 2, we first update the upper bounds of the minimum gap for optimal uniform WGFHSs.

\begin{theorem}\label{newbound}
Let $X$ be an optimal uniform WGFHS of period $ks$ over $F$ of size $s$ with minimum gap $e$. Then
$$
e\leq  \begin{cases}
             \lfloor \frac{s-1}{2} \rfloor -2, & \text {if} ~k=3, ~s\ge5, ~2\nmid s, \text{or} ~k\ge 4, ~s\ge k(k-1)+1, ~2\nmid s,\\
             \lfloor \frac{s-1}{2} \rfloor -1, & \text {otherwise}.
             \end{cases}
 $$
\end{theorem}

An optimal uniform WGFHS with parameters $(ks,s,k;e)$ achieving the upper bound in Theorem \ref{newbound} is called double optimal uniform WGFHS. In previous literatures, there have already been some results that are double optimal.

\begin{theorem}(\cite{LFY,LFM,LLZL})\label{k2odd}
There exists a double optimal uniform WGFHS with parameters (1) $(2s,s,2;\lfloor \frac{s-1}{2} \rfloor -1)$, where $s\ge 5$ is odd; (2) $(3s,s,3;\lfloor \frac{s-1}{2} \rfloor -2)$, where $s\ge 7$ and $s\equiv 1,5 \pmod 6$.
\end{theorem}

We give some new classes of double optimal uniform WGFHSs in Section 3, and completely finished the cases when $k=2$ and $k=3$, as shown in the theorem below.
Finally, Section 4 concludes the paper.

\begin{theorem}\label{k23}
(1) A double optimal uniform WGFHS with parameters $(2s,s,2;$ $\lfloor \frac{s-1}{2} \rfloor -1)$ exists if and only if $s\ge 5$;
(2) A double optimal uniform WGFHS with parameters $(3s,s,3;e)$ exists if and only if $s\ge6$, where $e=\lfloor\frac{s-1}{2}\rfloor-1$ if $2\mid s$, and $\lfloor\frac{s-1}{2}\rfloor-2$ otherwise.
\end{theorem}

\section{New Upper Bound of Minimum Gaps of Optimal Uniform WGFHSs}

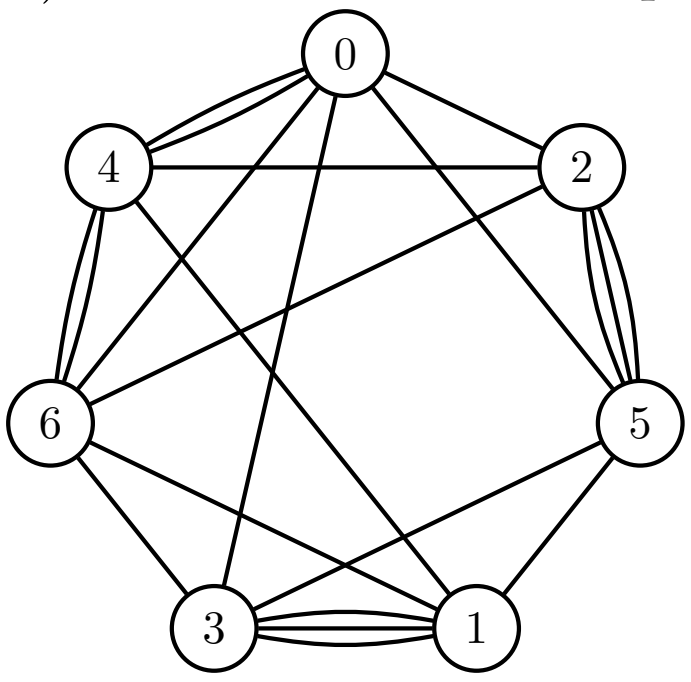
\begin{figure}[h]\centering
\vspace{-1.0em}
\caption{graph $G(X)$ of a WGFHS $X$ with parameters $(21,7,3;1)$}\label{graph}
\begin{tikzpicture}[
  scale=0.8,transform shape,
  vertex/.style={circle,draw=black,fill=white,line width=.8pt,
    minimum size=7mm,inner sep=0pt,font=\normalsize},
  every edge/.style={draw=black,line width=.8pt}
]
\foreach \v/\i in {0/0,2/1,5/2,1/3,3/4,6/5,4/6}
  \node[vertex] (v\v) at ({90-360*\i/7}:2.5cm) {$\v$};
\path (v1) edge (v6);
\path (v2) edge (v6);
\path (v0) edge (v2);
\path (v0) edge (v3);
\path (v3) edge (v5);
\path (v2) edge[bend left=-10] (v5);
\path (v2) edge (v5);
\path (v2) edge[bend left=10] (v5);
\path (v2) edge (v4);
\path (v0) edge[bend left=-5] (v4);
\path (v0) edge[bend left=5] (v4);
\path (v0) edge (v6);
\path (v4) edge[bend left=-5] (v6);
\path (v4) edge[bend left=5] (v6);
\path (v0) edge (v5);
\path (v1) edge (v5);
\path (v1) edge[bend left=-10] (v3);
\path (v1) edge (v3);
\path (v1) edge[bend left=10] (v3);
\path (v1) edge (v4);
\path (v3) edge (v6);
\end{tikzpicture}
\vspace{-1.0em}
\end{figure}

In this section, we will prove that if there exists an optimal uniform WGFHSs with parameters $(ks,s,k;e)$ where (1) $k=3,~ s\ge 5$ is odd, or (2) $k\ge 4, ~s\ge k(k-1)+1$ is odd, then $e\le \lfloor \frac{s-1}{2} \rfloor -2$.
By $a_0a_1\dots a_{m-1}$ we mean the path whose edges are $a_ia_{i+1}$ for $0\le i\le m-2$. By $a_0a_1\dots a_{m-1}a_0$ we mean the cycle obtained from the path above by adding the edge $a_{m-1}a_0$.
Let $X={(x_j)}_{j=0}^{v-1}$ be a WGFHS with parameters $(v,s,\lambda;e)$.
Since $\{x_i:0\le i\le v-1\}=\{f_i:0\le i\le s-1\}$,
treat each frequency $f_i, ~0\le i\le s-1$ as a vertex and $\{x_i,x_{i+1}\}, ~0\le i\le v-1$ as an undirected edge on the corresponding graph of $X$, denoted as $G(X)$ and allowing parallel edges.
For example, $X=(1,6,2,0,3,5,2,5,2,4,0,6,4,0,5,1,3,1,4,6,3)$ is an optimal uniform WGFHS with parameters $(21,7,3;1)$, then $G(X)$ is shown in Figure \ref{graph}.

Suppose $X$ is also optimal uniform, satisfying $k\ge3, ~s=2d+1, ~d\ge 2$ is an integer, and $e=d-1$. For distinct frequencies $a$ and $b$, define $N_{ab}=N_{ba}=|\{0\le i\le ks-1:\{x_i,x_{i+1}\}=\{a,b\}\}|$.
The degree of each vertex is $d(f_i)=2k$ for each frequency appear $k$ times and has exactly two neighbours on its both sides for each appearance in $X$.
We divide all possible edges of $G(X)$ into the following three types: (1) $\{a,a+d\}, ~0\le a\le d$; (2) $\{a,a+d+1\}, ~0\le a\le d-1$; (3) $\{a,a+d+r\}, ~(a,r)\in R, ~R=\{(a,r): r\ge 2, ~a\ge0, ~a+r\le d\}$.
Define a {\it basic cycle} $C=0,d+1,1,d+2,\dots,d-1,2d,d,0$. All edges of types (1) and (2) appear in this cycle.
Define the edge of type (3) a {\it chord}, and denote the number of its occurrences as $w_{a,r}$.
If there exists an integer $0\le i\le ks-1$ satisfying $x_{i-1}=x_{i+1}$, define it a {\it reversal centered at $x_{i}$}. Apparently, the total number of reversals in $X$ is exactly $H_{X,X}(2)\le k$.
Define $T=\sum_{(a,r)\in R} (r-1)w_{a,r}$.

\begin{lemma}\label{rev}
$H_{X,X}(2)\ge 2T$.
\end{lemma}
\begin{proof}
If $G(X)$ has no chords, then the underlying simple graph of $G(X)$ is $C$. So $T=0$, and $H_{X,X}(2)\ge 2T$ apparently.
Without loss of generality, suppose $G(X)=f_0f_1\dots f_{s-1}$ and $n_i$ be the edge-multiplicity of $f_if_{i+1}$.
Since $d(f_i)=2k$, we have $n_i+n_{i+1}=2k, 0\le i\le s-2$ and $n_{s-1}+n_{0}=2k$.
For $s-1$ is even, $n_0=n_2=\dots=n_{s-1}$, so $n_{s-1}+n_{0}=2n_{0}=2k$, which means $n_i=k$ for any $0\le i\le s-1$.
If $G(X)$ has chords, we can start with $C$, add chords, and adjust the edges on $C$. Suppose there is a chord $\{a,a+d+r\}$, then we can find an odd path $P=a,a+d+1,a+1,\dots,a+r-1,a+d+r$ on $C$. Figure \ref{path} shows the adjustment.
To keep the degree of each vertex, at either end-vertex of $P$, the subtraction is balanced by the newly added chord. At each internal vertex of $P$, the addition and the subtraction cancel.
Each time an actual chord appears corresponds to an adjustment.

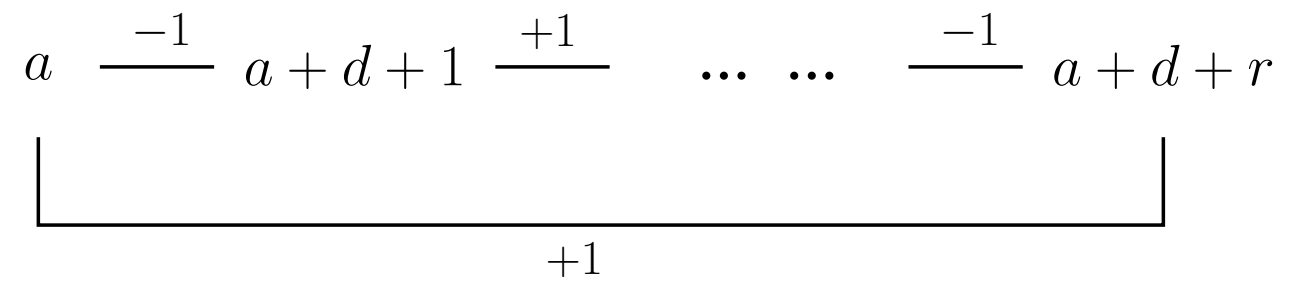
\begin{figure}[H]
\vspace{-1.0em}
\centering\caption{Adjustment path}\label{path}
\resizebox{0.6\linewidth}{!}{%
\begin{tikzpicture}[
  x=0.75pt,y=-0.75pt,
  every node/.style={
    inner sep=0pt,
    outer sep=0pt,
    font=\sffamily
  },
  line width=1.5pt,
  line cap=butt,
  line join=miter
]
\path[use as bounding box] (0,0) rectangle (800,200);

\node[font=\sffamily\bfseries\fontsize{33}{36}\selectfont]
  at (60,60) {$a$};
\node[font=\sffamily\bfseries\fontsize{33}{36}\selectfont]
  at (240,60) {$a+d+1$};
\node[font=\sffamily\bfseries\fontsize{33}{36}\selectfont]
  at (700,60) {$a+d+r$};
{
  \draw (95,60) -- (160,60);
  \draw (320,60) -- (385,60);
  \draw (555,60) -- (620,60);
  \foreach \dx in {-7,0,7}{
      circle[radius=2.0];
  }
}

\node[font=\sffamily\bfseries\fontsize{21}{23}\selectfont]
  at (130,40) {$-1$};
\node[font=\sffamily\bfseries\fontsize{21}{23}\selectfont]
  at (350,40) {$+1$};
\node[font=\sffamily\bfseries\fontsize{21}{23}\selectfont]
  at (590,40) {$-1$};

\foreach \xx in {440,450,460,490,500,510}{
  \fill (\xx,65) circle[radius=2.3];
}

\draw (60,100) -- (60,150)
  -- (700,150) -- (700,100);
\node[font=\sffamily\bfseries\fontsize{21}{23}\selectfont]
  at (365,170) {$+1$};

\end{tikzpicture}%
}
\vspace{-1.5em}
\end{figure}


Suppose the actual number of edge-multiplicity of $f_if_{i+1}$ in $G(X)$ is $N^{0}_i$, and the final number of edge-multiplicity of $f_if_{i+1}$ obtained by adjusting each chord only from odd paths on $C$ is $N^1_i$.
Apparently, $\sum_{(f_i,r)\in R}w_{f_i,r}+\sum_{(a,f_i-a-d)\in R}w_{a,f_i-a-d}+N^{0}_i+N^{0}_{i-1}=2k$ and $\sum_{(f_i,r)\in R}w_{f_i,r}+\sum_{(a,f_i-a-d)\in R}w_{a,f_i-a-d}+N^{1}_i+N^{1}_{i-1}=2k$.
Let $z_i=N^{1}_i-N^{0}_i$, then we have $z_{i-1}+z_i=0$ for $1\le i\le s-1$, and $z_{s-1}+z_0=0$.
Since $s-1$ is even, we have $z_0=z_2=\dots=z_{s-1}$, so $z_{s-1}+z_0=2z_0=0$ which means $z_i=0$ for any $0\le i\le s-1$. So $N^{01}_i=N^{1}_i$ for any $0\le i\le s-1$.
Then we just need to consider the case that each chord is adjusted only by odd paths.
Thus we can calculate that $N_{j,j+d}=k+\sum_{(a,r)\in R, ~a<j<a+r} w_{a,r}, 0\le j\le d$ and $N_{j,j+d+1}=k-\sum_{(a,r)\in R, ~a\le j<a+r} w_{a,r}, 0\le j\le d-1$.
Suppose $N_{uv}=k+q, ~q \ge 0$ and let $F_u$ be the number of occurrences of $u$ having $v$ on both sides in $X$.
Each of the $k$ occurrences of $u$ contributes at most one to $N_{uv}$, except that these $F_u$ occurrences contribute a second.
Thus $k + q \le k + F_u$, we have $F_u \ge q$.
The same argument gives at least $q$ reversals centered at $v$.
The excesses above $k$ of the edge $\{j, j + d\}$ sum to $\sum_{j=0}^{d} (N_{j,j+d}-k)= \sum _{(a,r)\in R} (r-1)w_{a,r}=T$.
Only $1 \le j \le d-1$ can have positive excess, these pairs are pairwise disjoint.
Their forced reversals therefore have different centers and can be summed without double counting. So the lower bound of reversals in $X$ is $H_{X,X}(2)\ge 2T$.
\end{proof}

\begin{lemma}\label{k3}
If there exists an optimal uniform WGFHSs with parameters $(3s,s,3; e)$ where $s\ge 5$ is odd, then $e\le \lfloor\frac{s-1}{2}\rfloor-2$.
\end{lemma}
\begin{proof}
By Lemma \ref{rev}, we have $2T\le H_{X,X}(2)\le 3$. So $T=0$ or $T=1$.
If $T=0$, the underlying simple graph of $G(X)$ is $C$, with each edge appears $3$ times.
The sequence $X$ moves along $C$.
A change of direction contributes exactly one reversal. Apparently, a cyclic sequence has an even number of changes, so there are either 0 or 2 direction changes.
With 0 changes, $X$ goes around the cycle ordering three times in one direction. Thus $H_{X,X}(s)=3s$.
With 2 changes, there are two constant-direction runs of lengths $p, q > 0$.
Then we have $p+q=3s, ~p-q\equiv 0 \pmod s$. So $\{p,q\}=\{s,2s\}$. Thus $H_{X,X}(s)\ge s$.
If $T=1$, there is exactly one chord $\{a, a+d+2\}$, and it occurs once. Suppose $U=a, ~V=a+d+1, ~W=a+1, ~Z=a+d+2$, then $P=UVWZ$.
After adjustment, we have $N_{UV}=2, ~N_{VW}=4, ~N_{WZ}=2, ~N_{ZU}=1$. Since $N_{VW}=4$ but each of $V,W$ occurs only 3 times, there are at least one reversal centered at $V$ and $W$ respectively.
Let $Q=r_0r_1\dots r_{s-3}$, where $r_0=U, ~r_{s-3}=Z$, be the complementary path of $P$ in $C$. 
We claim that no internal vertex $u = r_j, 1\le j\le s-4$ of $Q$ can be the center of a reversal in $X$.
Suppose $p = r_{j-1}$ and $q = r_{j+1}$, then $N_{pu} = N_{uq} = 3$.
Let $\alpha, \beta, \gamma$ count the occurrences of $u$ with neighbor patterns $(p, p), ~(q, q)$, and $(p, q)$ in $X$ respectively.
Then $2\alpha+\gamma = 3, ~2\beta+\gamma = 3$, which gives to $\alpha=\beta$.
If $\alpha=\beta\neq0$, together with those reversals centered at $Y,Z$, this would give $H_{X,X}(2) \ge 2+\alpha+\beta\ge 4$, a contradiction.
Therefore $X$ goes through $Q$ three times completely.
At least two of them have the same direction, so there exists the nonzero shift $\tau$ such that $H_{X,X}(\tau)\ge s-2$.
Both cases establish that $H(X)\ge s-2$. When $s\ge 7$, we have $H(X)\ge 5$, a contraction with $\lambda=3$. When $s=5$, we completely prove through exhaustive computer search that such sequence does not exist.
Thus, $e\le \lfloor\frac{s-1}{2}\rfloor-2$.
\end{proof}

\begin{lemma}\label{k4}
If there exists an optimal uniform WGFHSs with parameters $(ks,s,k;e)$ where $k\ge 4$ and $s\ge k(k-1)+1$ is odd, then $e\le \lfloor\frac{s-1}{2}\rfloor-2$.
\end{lemma}
\begin{proof}
By Lemma \ref{rev}, we have $2T\le H_{X,X}(2)\le k$, so $T\le \lfloor \frac{k}{2} \rfloor$.
Suppose there is a simple path of length $\ell$ on $C$, satisfying that each edge appears $k$ times, every internal vertex has only its two path neighbors as actual neighbors in $X$, and is never a reversal center.
Then $X$ goes through this path $k$ times completely.
At least two of them have the same direction, so there exists the nonzero shift $\tau$ such that $H_{X,X}(\tau)\ge \ell+1$. Since $H_{X,X}(\tau)\le k$, we have $\ell\le k-1$.
Now we divide $C$ into the simple paths.
If $T=0$, then the underlying simple graph of $G(X)$ is $C$, with each edge appears $k$ times.
If it has no reversals, then $H_{X,X}(s)=ks>k$. If it has reversals, split $C$ at every internal vertex that is a reversal center.
For example, if there exists a path $P=abcdef$ where $c$ is a reversal center, then we cut $P$ into $P_1=abc$ and $P_2=cdef$. Then $P_1$ and $P_2$ are simple paths.
There are at most $H_{X,X}(2)\le k$ such vertices, which means there are at most $k$ simple paths, so we have $s\le k(k-1)$, a contradiction.

If $T>0$, for each actual chord, we first delete its adjustment path $P=a,a+d+1,a+1,\dots,a+r-1,a+d+r$ form $C$. This path has $2r-1$ edges.
Let $U$ be the set of all these edges, and $q_0$ be the number of distinct chords.
Then we have $q_0 \le T$, and $|U|\le \sum_{w_{a,r}>0}(2r-1)\le 3\sum_{(a,r)\in R}(r-1)w_{a,r}=3T \le 3\lfloor \frac{k}{2} \rfloor <s$, which means the remaining edges in $C$ are nonempty. So the number of these edges is at least $s-3T$, and the edge-multiplicity of each edge is $k$. All these edges can form at most $q_0$ paths.
There are at least $2T$ reversal centered at vertices belonging to $U$, so there are at most $H_{X,X}(2)-2T$ reversals centered at internal vertices of these $q_0$ paths.
Cut each such path at every internal vertex that is a reversal center, and each cut adds one nonempty path.
The resulting number $M$ of these paths satisfies $M \le q_0+(H_{X,X}(2)-2T)\le k-T$.
All these paths satisfy the properties of the simple path mentioned above.
If $E$ denotes the number of remaining edges, then we have $s-3T\le E\le M(k-1) \le(k-T)(k-1)$.
But $s-3T-(k-T)(k-1)=s-k(k-1)+(k-4)T>0$, contradicting the preceding inequality. Thus, $e\le \lfloor\frac{s-1}{2}\rfloor-2$.
\end{proof}

{\noindent \underline{\it Proof of Theorem \ref{newbound}}}: Apply Lemmas \ref{k3}, and \ref{k4} to get the conclusion.
\qed

\section{New classes of double optimal uniform WGFHSs}

In this section, we will prove that there exists a double optimal uniform WGFHS for $k=2, ~s\ge 5$ and $k=3, ~s\ge 6$. Define $[a,b]=\{x\in \mathbb{Z}: a\le x\le b\}$, $[a,b]_i=\{x=a+ki: k\ge 0, a+ki\le b\}$. All the sets in the following tables are multiset and $2S$ denotes two copies of each element of $S$.

\begin{lemma}\label{k2s2}
There exists a double optimal uniform WGFHS with parameters $(2s,s,2;\lfloor \frac{s-1}{2} \rfloor$ $-1)$, where $s\ge 6$ and $s\equiv 2\pmod 4$.
\end{lemma}
\begin{proof}
Suppose $s=2t, ~t\ge 3$ is odd. Define $A_i=\langle (t-1)i\rangle_{2t}, ~B_i=\langle -(t-1)i\rangle_{2t}, ~C_i=\langle (t-1)i-1\rangle_{2t}, ~D_i=\langle -(t-1)i+1\rangle_{2t}, ~0\le i<t$.
Let $A={(A_i)}_{i=0}^{t-1}, ~B={(B_i)}_{i=0}^{t-1}, ~C={(C_i)}_{i=0}^{t-1}, ~D={(D_i)}_{i=0}^{t-1}$, then we will prove $X=(A,B,C,D)$ is a double optimal uniform WGFHS with parameters $(2s,s,2;\lfloor \frac{s-1}{2} \rfloor$ $-1)$.
Since $t$ is odd, then $gcd(t-1,2t)=2$. So $A, B$ run through $[0,s-2]_2$ once each, $C, D$ run through $[1,s-1]_2$ once each, which means each frequency appear $2$ times.
In each sequence, the inner difference is $\pm (t-1)$, so the absolute difference is $t-1$ or $t+1$.
Now we consider the junction of two sequences: $|A_{t-1}-B_0|=t+1, ~| B_{t-1}-C_0|=t, ~|C_{t-1}-D_0|=t-1, ~|D_{t-1}-A_0|=t$.
So we have $e=min| x_{i+1}-x_i|-1=t-2=\lfloor \frac{s-1}{2} \rfloor -1$.
For any frequency $f\in[0,s-2]_2$, it appears in $A$ and $B$ once each. Suppose $A_i=B_j=f, ~0\le i,j< t$, so $(t-1)i\equiv -(t-1)j\pmod {2t}$.
Since $gcd(t-1,2t)=2$, we have $i+j\equiv 0\pmod t$. Define $\delta=t+j-i$ is the difference of positions of $f$ in $X$.
Then we have $\delta=t$ when $i=0$, and $\delta=2t-2i$ when $1\le i<t$.
All position-difference of even frequency is $T=\{t\}\cup[2,2t-2]_2$.
Similarly, one can easily check that all position-difference of odd frequency is also $T$.
The position-difference of each frequency contributes one coincidence at $\tau=\delta$ and $\tau=4t-\delta$. And $T\cap(4t-T)=\emptyset$, so $H_{X,X}(\tau)=2$ when $\tau\in T\cup(4t-T)$, and $0$ otherwise. Then we have $H(X)=2$.
\end{proof}

\begin{lemma}\label{k2s4}
There exists a double optimal uniform WGFHS with parameters $(2s,s,2;\lfloor \frac{s-1}{2} \rfloor$ $-1)$, where $s\ge 8$ and $s\equiv 0\pmod 4$.
\end{lemma}
\begin{proof}
Suppose $s=2t, ~t\ge 4$ is even. Define $A_i=\langle (t-1)i\rangle_{2t}, ~B_i=\langle -(t-1)i\rangle_{2t}, ~0\le i<2t$. Let $A^1={(A_i)}_{i=0}^{t-2}, ~B^1=(B_0, B_1), ~A^2={(A_i)}_{i=t-1}^{2t-1}, ~B^2={(B_i)}_{i=2}^{2t-1}$, then we will prove $X=(A^1,B^1,A^2,B^2)$ is a double optimal uniform WGFHS with parameters $(2s,s,2;\lfloor \frac{s-1}{2} \rfloor$ $-1)$.
Since $t$ is even, then $gcd(t-1,2t)=1$. So $A, B$ run through $[0,s-1]$ once each, which means each frequency appear $2$ times.
In each sequence, the inner difference is $\pm (t-1)$, so the absolute difference is $t-1$ or $t+1$. Now we consider the junction of two sequences: $|A_{t-2}-B_0|=t+2, ~|B_1-A_{t-1}|=t, ~| A_{2t-1}-B_2|=t-1, ~|B_{2t-1}-A_0|=t-1$.
So we have $e=min| x_{i+1}-x_i|-1=t-2=\lfloor \frac{s-1}{2} \rfloor -1$.
For any frequency $f$, it appears in $A$ and $B$ once each.
Suppose $A_i=B_j, ~0\le i,j<2t$, so
$i+j \equiv 0\pmod {2t}$. Suppose $x_u=A_i, ~x_v=B_j$, define $T=\{|u-v|:x_u=x_v, 0\le u\neq v<4t\}$. Then we have
$$
\mid u-v\mid =  \begin{cases}
             t-1, & \text {if} ~i=0,\\
            4t-2i, & \text {if} ~1\le i\le t-2,\\
            4t-2-2i, & \text {if} ~t-1\le i\le 2t-2,\\
            t+1, & \text {if} ~i=2t-1.
            \end{cases}
$$
Thus $T=\{t-1,t+1\}\cup[2t+4,4t-2]_2\cup[2,2t]_2$, and $4t-T=\{3t+1,3t-1\}\cup[2,2t-4]_2\cup[2t,4t-2]_2$. So $H_{X,X}(\tau)=2$ when $\tau\in [2,2t-4]_2\cup\{2t\}\cup[2t+4,4t-2]_2$, and $1$ when $\tau\in\{t\pm1,3t\pm1,2t-2,2t+2\}$, and $0$ otherwise. Then we have $H(X)=2$.
\end{proof}

\begin{lemma}\label{k3s6}
There exists a double optimal uniform WGFHS with parameters $(3s,s,3;\lfloor \frac{s-1}{2} \rfloor$ $-1)$, where $s\ge 6$ and $s\equiv 2\pmod 4$.
\end{lemma}
\begin{proof}
Suppose $s=2t, ~t\ge 3$ is odd. Define $A_i=\langle (t-1)(1+i)-2\rangle_{2t}, ~B_i=\langle (t-1)(1-i)\rangle_{2t}, ~C_i=\langle (t-1)i+1\rangle_{2t}, ~D_i=\langle -(t-1)i+1\rangle_{2t}$, and  $E_{2i}=i, ~E_{2i+1}=t+i, ~0\le i<t$.
Let $A={(A_i)}_{i=0}^{t-1}, ~B={(B_i)}_{i=0}^{t-1}, ~C={(C_i)}_{i=0}^{t-1}, ~D={(D_i)}_{i=0}^{t-1}$ and $E={(E_i)}_{i=0}^{2t-1}$, then we will prove $X=(A,B,C,D,E)$ is a double optimal uniform WGFHS with parameters $(3s,s,3;\lfloor \frac{s-1}{2} \rfloor$ $-1)$.
Since $t$ is odd, then $gcd(t-1,2t)=2$. So $A, B$ run through $[0,s-2]_2$ once each, $C, D$ run through $[1,s-1]_2$ once each, and $E$ contain $[0,s-1]$ once each, which means each frequency appear $3$ times.
The inner absolute difference is $t-1$ or $t+1$ in $A,B,C,D$, and $t$ or $t-1$ in $E$. Now we consider the junction of two sequences: $| A_{t-1}-B_0|=t-1, ~|B_{t-1}-C_0|=2t-3, ~|C_{t-1}-D_0|=t+1, ~| D_{t-1}-E_0|=t, ~|E_{2t-1}-A_0|=t+2$.
So we have $e=min|x_{i+1}-x_i|-1=t-2=\lfloor \frac{s-1}{2} \rfloor -1$.
For any frequency $f$, define $a_f, b_f, c_f, d_f$ and $e_f$ are its positions. Actual positions of $f$ in $X$ can be seen in Table \ref{table5}.

\begin{table}[h]\centering
\vspace{-1.0em}
\caption{actual positions of frequency $f$ in $X$}\label{table5}
\begin{tabular}{|c|c|c|c|c|c|}
  \hline
   & position & $f$ &  & position & $f$ \\
  \hline
        & $t-3-f$ & $0\le f<t-1, ~2\mid f$ &       & $2t$ & $f=1$ \\
  $a_f$ & $2t-3-f$ & $t-1\le f \le 2t-4, ~2\mid f$ &   $c_f$    & $3t+1-f$ & $3\le f\le t, ~2\nmid f$ \\
  & $t-1$ & $f=2t-2$ & & $4t+1-f$ & $t<f\le 2t-1, ~2\nmid f$ \\
  \hline
  $b_f$   & $t+1+f$ & $0\le f< t-1, ~2\mid f$ &   $d_f$    & $3t-1+f$ & $1\le f\le t, ~2\nmid f$ \\
        & $1+f$ & $t-1\le f\le 2t-2, ~2\mid f$ &       & $2t-1+f$ & $t<f\le 2t-1, ~2\nmid f$ \\
  \hline
  $e_f$ & $4t+2f$ & $0\le f<t$ & $e_f$ & $2t+2f+1$ & $t\le f\le 2t-1$ \\
  \hline
\end{tabular}
\end{table}
Calculating the position-difference for all possible pair types, we obtain Table \ref{table1} below.
Define $T=\{2,t,t+2,3t-1\}\cup(([6,3t-3]_6 \cup [8,3t-7]_6)\setminus \{2t\}) \cup [t+6,2t-1]_6 \cup [t+8,2t-1]_6$. Apparently, $T\cap (6t-T)=\emptyset$. Then we have $H_{X,X}(\tau)=3$ when $\tau \in T\cup (6t-T)$, and $<3$ otherwise. So $H(X)=3$.
\end{proof}

\begin{table}[h]\centering
\vspace{-2.5em}
\footnotesize
\caption{pair type and $\tau$ values for $s\ge 6$ and $s\equiv 2\pmod 4$}\label{table1}
\begin{tabular}{|c|c|}
  \hline
  pair type & $\tau$ values \\
  \hline
  (A,B) & $[2,2t-2]_2 \cup [4t+2,6t-2]_2 \cup \{t,5t\}$  \\
  (A,E) & $[6,3t-3]_6 \cup [8,3t-7]_6 \cup [3t+3,6t-6]_6 \cup [3t+7,6t-8]_6 \cup\{t,t+2,5t,5t-2\}$ \\
  (B,E) & $2[2t+4,3t-1]_2 \cup 2[3t+1,4t-4]_2 \cup \{t+2, 2t+2,3t-1,3t+1,4t-2,5t-2\}$ \\
  (C,D) & $[2,2t-2]_2 \cup [4t+2,6t-2]_2 \cup \{t,5t\}$ \\
  (C,E) & $[t+6,4t-3]_6 \cup [t+8,4t-7]_6 \cup [2t+3,5t-6]_6 \cup [2t+7,5t-8]_6 \cup\{2t,2t+2,4t,4t-2\}$ \\
  (D,E) & $2[t+4,2t-1]_2 \cup 2[4t+1,5t-4]_2 \cup \{2, t+2,2t+1,4t-1,5t-2,6t-2\}$ \\
  \hline
\end{tabular}
\end{table}

\begin{lemma}\label{k3s8}
There exists a double optimal uniform WGFHS with parameters $(3s,s,3;\lfloor \frac{s-1}{2} \rfloor$ $-1)$, where $s\ge 8$ and $s\equiv 0\pmod 4$.
\end{lemma}
\begin{proof}
Suppose $s=2t, t\ge 4$ is even. Define $A_i=\langle (t-1)(1+i)\rangle_{2t}, ~B_i=\langle 1-(t-1)i\rangle_{2t}, ~0\le i<2t$, and $C_{2i}=i, ~C_{2i+1}=t+i, ~0\le i<t$.
Let $X = X_j$ for $j=\langle t\rangle_{6}\in \{0,2,4\}$, define

\noindent $X_0=({(A_i)}_{i=0}^{t-1},{(B_i)}_{i=0}^{t-2},{(C_i)}_{i=0}^{2t-1},{(A_i)}_{i=t+1}^{2t-1},{(B_i)}_{i=t}^{2t-1},B_{t-1},A_{t}),$\\
\noindent
$X_2=({(A_i)}_{i=0}^{t-2},{(B_i)}_{i=1}^{t-2},{(C_i)}_{i=0}^{2t-1},{(A_i)}_{i=t+1}^{2t-1},{(B_i)}_{i=t}^{2t-1},B_{t-1},A_{t-1},B_{0},A_{t}),$\\
\noindent
$X_4=({(A_i)}_{i=0}^{t-3},{(B_i)}_{i=0}^{t-2},{(C_i)}_{i=0}^{2t-1},{(A_i)}_{i=t+1}^{2t-2},A_{t-2},A_{t-1},A_{2t-1},{(B_i)}_{i=t}^{2t-1},B_{t-1},A_{t}),$

\noindent then we will prove $X$ is a double optimal uniform WGFHS with parameters $(3s,s,3;\lfloor \frac{s-1}{2} \rfloor$ $-1)$. When $j=0$,
since $t$ is even, then $gcd(t-1,2t)=1$. So $A, B$ and $C$ run through $[0,s-1]$ once each, which means which means each frequency appear $3$ times.
The inner absolute difference of two constant positions is $t-1$ or $t+1$ in $A,B$, and $t$ or $t-1$ in $C$. Now we consider the junction of two sequences: $|A_{t-1}-B_0|=t-1, ~|B_{t-2}-C_0|=t-1, ~|C_{2t-1}-A_{t+1}|=t+1, ~|A_{2t-1}-B_t|=t+1, ~| B_{2t-1}-B_{t-1}|=t, ~|B_{t-1}-A_{t}|=2t-1, ~|A_{t}-A_0|=t$.
So we have $e=min|x_{i+1}-x_i|-1=t-2=\lfloor \frac{s-1}{2} \rfloor -1$.
When $j=2$, compared with the case $j=0$, we also need to consider $|A_{t-2}-B_1|=t+1, ~|B_{t-1}-A_{t-1}|=t, ~|A_{t-1}-B_0|=t-1, ~|B_0-A_t|=2t-2$. When $j=4$, $|A_{t-3}-B_0|=t+1, ~|A_{2t-2}-A_{t-2}|=t, ~|A_{t-2}-A_{t-1}|=t-1, ~|A_{t-1}-A_{2t-1}|=t$. So these two cases still satisfy $e=min|x_{i+1}-x_i|-1=t-2=\lfloor \frac{s-1}{2} \rfloor -1$.
For any frequency $f$, define $a_f, b_f$ and $c_f$ are its three positions. Actual positions of $f$ in $X_0$ can be seen in Table \ref{table3}. The remaining two cases can be checked directly, and we omit the details here.

\begin{table}[h]\centering
\vspace{-1.0em}
\caption{actual positions of frequency $f$ in $X_0$}\label{table3}
\begin{tabular}{|c|c|c|c|c|c|}
  \hline
   & position & $f$ &  & position & $f$ \\
  \hline
        & $5t-3-f$ & $0\le f<t, ~2\mid f$ &       & $6t-2$ & $f=0$ \\
        & $2t-1-f$ & $t\le f \le 2t-2, ~2\mid f$ &       & $5t-3+f$ & $2\le f\le t, ~2\mid f$ \\
  $a_f$ & $t-1-f$ & $1\le f<t, ~2\nmid f$ & $b_f$ & $f-1$ & $t<f\le 2t-2, ~2\mid f$ \\
        & $6t-3-f$ & $t< f\le 2t-3, ~2\nmid f$ &       & $t-1+f$ & $1\le f<t, ~2\nmid f$ \\
        & $6t-1$ & $f=2t-1$ &       & $4t-3+f$ & $t<f\le 2t-1, ~2\nmid f$ \\
  \hline
  $c_f$ & $2t-1+2f$ & $0\le f<t$ & $c_f$ & $2f$ & $t\le f\le 2t-1$ \\
  \hline
\end{tabular}
\end{table}
For each pair type of equal positions, list either its positive
difference $\delta$ or its complement $6t-\delta$, see Table \ref{table2}.
Define $T=\{t+2,2t+3\} \cup [4,3t-8]_6 \cup [6,3t-6]_6 \cup [t+1,2t+1]_6 \cup [t+3,2t-3]_6$. Apparently, $T\cap (6t-T)=\emptyset$. Then we have $H_{X,X}(\tau)=3$ when $\tau \in T\cup (6t-T)$, and $<3$ otherwise. So $H(X)=3$.
When $j=2$, let $T=\{2,t,2t+1,2t+3\}\cup [4,3t-8]_6 \cup [6,3t-6]_6 \cup [t+5, 2t-3]_6 \cup [t+7, 2t-1]_6$. When $j=4$, let $T=((\{3t-4\} \cup [4,3t-2]_6 \cup [6,3t-6]_6)\setminus \{2t-2,2t\})\cup \{t+1,2t-1,2t+3\}\cup[t+5,2t-5]_6 \cup[t+7, 2t-3]_6$. For these two cases, we also have $H_{X,X}(\tau)=3$ when $\tau \in T\cup (6t-T)$, and $<3$ otherwise. So $H(X)=3$.
\end{proof}

\begin{table}[h]\centering
\vspace{-2.0em}
\caption{pair type and $\delta$ or $6t-\delta$ values for $s\ge 8, ~s\equiv 0\pmod 4$ and $j=0,2,4$}\label{table2}
\begin{tabular}{|c|c|c|}
  \hline
  j & pair type & $\delta$ or $6t-\delta$ values \\
  \hline
   & (A,B) & $2[2,2t-4]_2 \cup \{3,t+1,t+2,2t-2\}$  \\
  0 & (A,C) & $[4,3t-2]_6 \cup [6,3t-6]_6 \cup [t+1,4t-5]_6 \cup [t+3,4t-3]_6 \cup\{2t+1\}$ \\
   & (B,C) & $2[2t,3t-4]_2 \cup 2[t+3,2t-1]_2 \cup \{t+1,2t-2,2t+1,2t+3\}$ \\
  \hline
     & (A,B) & $2[2,2t-4]_2 \cup \{2,5,t,t+1\}$  \\
  2 & (A,C) & $[4,3t-2]_6 \cup [6,3t-6]_6 \cup [t+5,4t-7]_6 \cup [t+1,4t-5]_6 \cup\{2t+1,2t+3\}$ \\
   & (B,C) & $2[2t,3t-4]_2 \cup 2[t+3,2t-1]_2 \cup \{2t-2, 2t+3\} \cup 2\{2t+1\}$ \\
  \hline
   & (A,B) & $[2,2t-2]_2 \cup [4,2t-4]_2 \cup 2\{t+1\} \cup \{3,2t+3\}$  \\
  4 & (A,C) & $[4,3t-2]_6 \cup [6,3t-6]_6 \cup [t+5,4t-7]_6 \cup [t+7,4t-5]_6 \cup\{2t+3,3t-4,3t\}$ \\
   & (B,C) & $2[2t+2,3t-2]_2 \cup 2[t+3,2t-1]_2 \cup \{t+1, 2t-1,2t,2t+1\}$ \\
  \hline
\end{tabular}
\end{table}

\begin{lemma}\label{s3q}
There exists a double optimal uniform WGFHS with parameters $(3s,s,3;\lfloor\frac{s-1}{2}\rfloor$ $-2)$, where $s\ge 9$ and $s\equiv 3\pmod 6$.
\end{lemma}
\begin{proof}
Suppose $s=3q$ where $q\ge 3$ is odd. Let $a=\frac{s-1}{2}$.
Define $A_i=\langle 1+ai\rangle_{s}, ~E_i=\langle 1-ai\rangle_{s}, ~0\le i<s$, and $B_j=\langle 2+(a-1)j\rangle_{s}, ~C_j=\langle (a-1)j\rangle_{s}, ~D_j=\langle 1-(a-1)j\rangle_{s}, ~0\le j<q$.
Let $A={(A_i)}_{i=0}^{s-1}, ~B={(B_i)}_{i=0}^{q-1}, ~C={(C_i)}_{i=0}^{q-1}, ~D={(D_i)}_{i=0}^{q-1}, ~E={(E_i)}_{i=0}^{s-1}$, then we will prove $X=(A,B,C,D,E)$ is a double optimal WGFHS with parameters $(3s,s,3;\lfloor \frac{s-1}{2} \rfloor$ $-2)$.
Since $gcd(a,s)=1$ and $gcd(a-1,s)=3$, So $A, E$ run through $[0,s-1]$ once each, $B, C$ and $D$ contain all frequencies congruent to $2,0,1$ modulo $3$ once each respectively, which means each frequency appear $3$ times.
The inner absolute difference is $a$ or $a+1$ in $A$ and $E$, and $a-1$ or $a+2$ in $B,C$ and $D$.
Now we consider the junction of two sequences: $|A_{s-1}-B_0|=a, ~| B_{q-1}-C_0|=a+4, ~|C_{q-1}-D_0|=a+1, ~|D_{q-1}-E_0|=a-1, ~| E_{s-1}-A_0|=a$. So we have $e=min|x_{i+1}-x_i|-1=a-2=\lfloor \frac{s-1}{2} \rfloor -2$.
For any frequency $f$, define $a_f, b_f, c_f, d_f$ and $e_f$ are its positions. Actual positions of $f$ in $X$ can be seen in Table \ref{table6}.

\begin{table}[h]\centering
\vspace{-1.0em}
\caption{actual positions of frequency $f$ in $X$}\label{table6}
\begin{tabular}{|c|c|c|c|c|c|}
  \hline
   & position & $f$ &  & position & $u$ \\
  \hline
        & $2$ & $f=0$ &    & $3q$ & $u=0$   \\
  $a_f$ & $0$ & $f=1$ &   $b_f,$ & $4q-2u$ & $1\le u\le \frac{q-1}{2}$ \\
  & $3q+2-2f$ & $2\le f\le \frac{3q+1}{2}$ & $f=3u+2$  & $5q-2u$ & $\frac{q+1}{2}\le u\le q-1$  \\
   & $6q+2-2f$ & $\frac{3q+3}{2}\le f\le 3q-1$ & & & \\
  \hline
  & $9q-2$ & $f=0$ &  & $4q$ & $u=0$ \\
   $e_f$   & $6q-2+2f$ & $1\le f\le \frac{3q+1}{2}$ & $c_f,~f=3u$ &  $5q-2u$ & $1\le u\le \frac{q-1}{2}$\\
   & $3q-2+2f$ & $\frac{3q+3}{2}\le f\le 3q-1$ & &  $6q-2u$ & $\frac{q+1}{2}\le u\le q-1$ \\
   \hline
   & & & $d_f,$ & $5q+2u$ & $0\le u\le \frac{q-1}{2}$\\
   & & & $f=3u+1$ & $4q+2u$ & $\frac{q+1}{2}\le u\le q-1$ \\
   \hline
\end{tabular}
\end{table}

Each pair of equal-frequency positions contributes one match at each of two complementary shifts $\tau$ and $9q-\tau$.
Since the sequence length is odd, so exactly one of these two shifts is even. Count that even shift as $2r$, where $1\le r\le(9q-1)/2$.
For each pair type, list its corresponding integer $r$ for even shifts, see Table \ref{table4}.
Define $T=[2q+4,4q]_2 \cup [4q+2,8q-6]_4 \cup\{6q-2\}$. Apparently, $T\cap (9q-T)=\emptyset$. Then we have $H_{X,X}(\tau)=3$ when $\tau\in T\cup (9q-T)$, and $<3$ otherwise. So $H(X)= 3$.
\end{proof}

\begin{table}[h]\centering
\vspace{-2.0em}
\caption{pair type and $r$ values for $s\ge 9$ and $s\equiv 3\pmod 6$}\label{table4}
\begin{tabular}{|c|c|c|c|}
  \hline
  pair type & $r$ values & pair type & $r$ values \\
  \hline
  (A,E) & $[1,3q]$ & (B,E) & $\{3q-1\}\cup[q+3,3q-1]_2$ \\
  (A,B) & $\{1\}\cup[3q,4q-2]$ & (C,E) & $[2q+1,4q-1]_2$ \\
  (A,C) & $[q,2q-1]$ & (D,E) & $[3q+1,4q]$ \\
  (A,D) & $\{2q\}\cup[q+2,3q-2]_2$ &   &   \\
  \hline
\end{tabular}
\end{table}

{\noindent \underline{\it Proof of Theorem \ref{k23}}}:

\noindent (1) Apply Theorem \ref{k2odd} (1), and Lemmas \ref{k2s2}, \ref{k2s4} to get the conclusion;

\noindent (2) By Theorem \ref{k2odd} (2) and Lemmas \ref{k3s6}, \ref{k3s8} and \ref{s3q}, we can get the conclusion.
\qed

\section{Conclusion}

This paper updates the upper bounds on the minimum gap $e$ of optimal uniform WGFHSs with parameters $(ks,s,k;e)$ where (1) $k=3, s\ge5$ is odd, (2) $k\ge 4$ and $s\ge k(k-1)+1$ is odd. For $k=2$ and $k=3$, we completely prove the existence of double optimal uniform WGFHSs with $s\ge5$ and $s\ge 6$ respectively. Furthermore, we can prove that if there exists an optimal uniform WGFHSs where $k\in\{4,5\}$ and $7\le s\le k(k-1)-1, 2\nmid s$, then $e\le \lfloor\frac{s-1}{2}\rfloor-2$ by exhaustive computer search. So we propose the following conjecture.

\begin{conjecture}
Let $X$ be an optimal uniform WGFHS of period $ks$ over $F$ of size $s$ with minimum gap $e$. Then
$$
e\leq  \begin{cases}
             \lfloor \frac{s-1}{2} \rfloor -2, & \text {if} ~k\ge 3, ~\text{and} ~2\nmid s,\\
             \lfloor \frac{s-1}{2} \rfloor -1, & \text {otherwise}.
             \end{cases}
 $$
\end{conjecture}

Also, we obtain some examples of optimal uniform WGFHSs for $k=4, s\in\{6,12,16\}\cup [7,25]_2$ and $k=5, s\in\{6,23\}\cup [7,19]_2$ achieving the upper bound of $e$ in the above conjecture, which will not be listed in this paper, thus we propose the conjecture as follows.

\begin{conjecture}
A double optimal uniform WGFHSs with parameters $(ks,s,k;e)$ exists if and only if $k=2, s\ge5$ or $k\ge 3, s\ge 6$.
\end{conjecture}

\noindent It would be interesting to give constructions of double optimal uniform WGFHSs. Also, studying the double optimality in non-uniform cases is pretty meaningful.
We leave these two problems as the prospective directions for further investigation.

\noindent{\bf Acknowledgments:} This work was supported by the National Natural Science Foundation of China
under Grant No. 12471313 (Haitao Cao).

\end{document}